\documentclass[twocolumn]{aastex63}
\usepackage{graphicx}
\graphicspath{{images/}}
\usepackage[caption2]{ccaption}
\usepackage{natbib}
\usepackage{CJK}

\begin{document}
\begin{CJK*}{UTF8}{gbsn}
\title{A Complete 16 $\mu$m selected Galaxy Sample at $z \sim 1$. II: Morphological Analysis}

\author[0000-0001-9143-3781]{Piaoran Liang}
\affiliation{Chinese Academy of Sciences South America Center for Astronomy (CASSACA), National Astronomical Observatories(NAOC),
20A Datun Road, Beijing 100012, China}
\affiliation{School of Astronomy and Space Science, University of Chinese Academy of Sciences, Beijing 101408, China}

\author[0000-0002-7928-416X]{Y. Sophia Dai(戴昱)}
\affiliation{Chinese Academy of Sciences South America Center for Astronomy (CASSACA), National Astronomical Observatories(NAOC),
20A Datun Road, Beijing 100012, China}

\author[0000-0001-6511-8745]{Jia-Sheng Huang}
\affiliation{Chinese Academy of Sciences South America Center for Astronomy (CASSACA), National Astronomical Observatories(NAOC),
20A Datun Road, Beijing 100012, China}
\affiliation{Center for Astrophysics \textbar\ Harvard \& Smithsonian, 60 Garden St., Cambridge, MA 02138 USA}

\author[0000-0003-0202-0534]{Cheng Cheng}
\affiliation{Chinese Academy of Sciences South America Center for Astronomy (CASSACA), National Astronomical Observatories(NAOC),
20A Datun Road, Beijing 100012, China}

\author[0000-0001-9471-831X]{Yaru, Shi}
\affiliation{Chinese Academy of Sciences South America Center for Astronomy (CASSACA), National Astronomical Observatories(NAOC),
20A Datun Road, Beijing 100012, China}
\affiliation{School of Astronomy and Space Science, University of Chinese Academy of Sciences, Beijing 101408, China}
\correspondingauthor{Y. Sophia Dai}
\email{ydai@nao.cas.cn}

\begin{abstract}
We present morphological analysis of the 16$\mu$m flux-density-limited galaxy sample
at 0.8$<z<$1.3 from \citet{huang21}.
At the targeted redshift, the 16$\mu$m emission corresponds to the 
Polycyclic aromatic hydrocarbon (PAH) feature from intense star formation, 
or dust heated by AGN (Active galactic nuclei). 
Our sample of 479 galaxies are dominated by Luminous Infrared Galaxies (LIRGs, 67\%)
in three CANDLES fields (EGS, GOODS-N, and GOODS-S), 
and are further divided into AGN dominated, star-forming dominated, composite,  
and blue compact galaxies by their 
spectral energy distribution (SED) types. 
The majority of our sample (71\%) have disky morphologies, 
with the few AGN dominated galaxies being more bulge-dominanted than the star-forming dominated and composite galaxies. 
The distribution of our sample on the Gini vs. M$_{\text{20}}$ plane
is consistent with previous studies, 
where the S{\'e}rsic index $n$ shows an increasing trend towards the smaller M$_{\text{20}}$ and higher Gini region
below the dividing line for mergers.
The subsample of ULIRGs follow a steep size-mass relation 
that is closer to the early-type galaxies. 
In addition, as the 4.5 $\mu$m luminosity excess ($L_{4.5}^{Exc}$, proxy for AGN strength) increases, 
our sample appear to be more bulge-dominated (i.e. higher $n$). 
Based on the sSFR and compactness ($log_{10}\Sigma_{1.5}, \Sigma_{1.5}=M_*/R_e^{1.5}$) diagram,
the majority of our LIRG-dominated galaxy sample follow a secular evolution track, 
and their distribution can be explained without involving any merging activities. 
Out of the 16 ULIRGs in our sample, six are compact with strong AGN contributions,
likely evolving along the fast-track from more violent activities. 
\end{abstract}


\section{introduction}
\defcitealias{huang21}{H21}
\defcitealias{van14}{V14}
\defcitealias{fang18candels}{F18}
\defcitealias{barro13}{B13}

Luminous Infrared Galaxies (LIRGs, $L_{IR}>10^{11}L_{\odot}$) play an important role in the mass assembly at the cosmic noon. 
In the early space astronomy era, 
galaxies with intensive star formation were only found in infrared observations such as with Infrared Astronomical Satellite (\textit{IRAS}) and the Infrared Space Observatory (\textit{ISO}). 
They are rather rare in the local universe \citep{soifer91, sanders96}. 
Both \textit{Spitzer} and \textit{Herschel} space telescopes made it possible to perform large infrared galaxy surveys \citep[e.g.][]{dole04firspitzer,lefloch05,armus09goals,magnelli13herschel} and had substantial detection of infrared galaxies at intermediate and high redshifts. 
These surveys also found that LIRGs had a dominant contribution to star-forming activities in the universe at $z \sim 1$ \citep{lefloch05}. 
Understanding the properties and evolution of these LIRGs is therefore critical in understanding galaxy evolution at the cosmic noon.

In the mid-infrared, the selection of LIRG samples is subject to complicated biases. 
\textit{ISO} and \textit{Spitzer} spectroscopy in the mid-infrared showed that a galaxy can have several predominant spectral features in its mid-infrared SED (spectral energy distribution), including the polycyclic aromatic hydrocarbon (PAH) emission features, strong continuum, or silicate absorption/emission\citep[e.g.][]{farrah08, huang09}. 
Galaxy populations in a mid-infrared flux-limited sample are strongly redshift-dependent, 
as the spectral features are shifted to certain selected bands. 
A similar bias also occurs in the far-infrared selected samples, 
if selected in the same band, 
galaxies with higher redshifts will be biased towards high dust temperatures \citep[e.g.][]{magdis10,chapman10,magnelli10firTd}.

Regardless of the redshift bias, one of the most effective methods 
to select galaxy samples is by their spectral features. 
\citet{huang07} studied an 8$\mu$m selected local galaxy sample and found that it consists of predominantly star-forming galaxies with strong PAH emission features. 
This method was also used to select star-forming galaxies with strong PAH emission features at z$\sim$1 and 2 in the 16 and 24$\mu$m bands \citep[e.g.][]{huang09, fang14, huang21}. 
\citet{magdis10} studied the far-infrared properties of a PAH selected galaxy sample at z$\sim$2, and found no bias in dust temperatures. 
On the other hand,  AGN (active galactic nuclei) fraction in PAH-selected galaxy samples increases from 2\% locally \citep{huang07},  to 15\% at z $\sim$ 1 \citep[hereafter H21]{huang21}, and 25\% at z $\sim$ 2 \citep[e.g.][]{huang09,fang14}. 
This is not due to cosmic evolution, but rather a luminosity effect. 
Most galaxies selected at z$\sim$1 are LIRGs, and are Ultra-Luminous Infrared Galaxies (ULIRGs, $L_{IR}>10^{12}L_{\odot}$) at z$\sim$2.
\citetalias{huang21} carried out
a mid-infrared SED analysis for a complete 16μm selected sample, mainly divided the galaxies into four
types (AGN, star-forming, composite, and blue compact galaxies), and quantified the AGN strength with
the 4.5$\micron$m luminosity excess ($L_{4.5}^{Exc}$) in each galaxy.

Morphologies of (U)LIRGs are critical in studying their formation and 
to identify the main power source of their observed high luminosities. 
Previous studies \citep[e.g.][]{sanders96, farrah01ulirgmerge, veilleux02} on the morphology of local (U)LIRGs found that more than 90\% of the (U)LIRG population appear to be merging systems, and argued that merging can effectively funnel the gas into the central part of galaxies and induce intensive star formation along the way. 
Local LIRG survey such as GOALS (Great Observatories All-sky LIRG Survey, \citet{sanders03}) 
found that mergers are more dominant in the more luminous LIRGs,
while less luminous LIRGs are dominated by minor mergers or consistent with secular evolution, 
with no significant morphological disturbance \citep{larson16}. 
The merger fraction seems to decline at high redshifts \citep[e.g.][]{bell05, melbourne05, kartaltepe12}. 
Possibly, galaxies at high redshifts have enough gas to undergo very intensive star formation without involving galaxy merging \citep[e.g.][]{daddi10, tacconi10, shapley11review, carilli13gasreview}.
Morphological studies on infrared selected galaxies at high redshifts are difficult
as infrared selected galaxies such as (U)LIRGs are also often subject to heavy dust extinction. 
At z $\approx$ 1, most of the optical imaging detects photons from a galaxy at rest-frame ultraviolet or wavelengths shorter than 4000\AA, whose morphology can be easily altered by the dust lane \citep{Meurer99, madau14SFH}. 
In addition, resolution limit has 
also hindered the morphological studies of $z\sim$1 galaxies,
until
the \textit{HST} (Hubble Space Telescope) WFC3 camera  \citep{kartaltepe12} and surveys such as the Cosmic Assembly Near-Infrared Dark Energy Legacy Survey (CANDELS; \citet[e.g.][]{grogin11candels, koekemoer11candels}) became available.
For instance, using HST optical imaging data \citep[e.g.][]{bell05, melbourne05, elbaz07, melbourne08} 
found that the majority of LIRGs at z $>$ 0.5 have spiral morphologies. 
While in \citet{kartaltepe12}, the majority of ULIRGs display merger like morphologies
with distorted and irregular features. 
Recent JWST (James Webb Space Telescope) observations have also revealed that previously considered isolated
ULIRGs are in fact merging systems, with a heavily obscured second nucleus buried
in the galaxy center \citep{huang23}. 

Numerous studies have explored the relationship between the galaxy morphologies and SFRs, 
such as the connection between interacting systems and elevated SFRs \citep[e.g.][]{larson78, barton00, lin07, shah22}, 
and the scenario of morphology quenching of the star formation \citep[e.g.][]{martig09, guo15}. 
The collisions of gas generated by interacting galaxies can lead to rapid star formation. 
There are substantial evidences that confirm the increased star formation activity in merger systems \citep[e.g.][]{larson78, barton00, lin07, shah22}, while mergers can also result in the formation of a bulge \citep{aguerri01}. 
Studies have shown that nearly all nearby ULIRGs are interacting merger systems, 
revealing a substantial correlation between mergers and star formation activities \citep[e.g.][]{sanders88, sanders96, huang23}. 
However, in the high redshift universe, things became complicated. 
Some studies showed that the fraction of merger systems in ULIRGs,
as well as the contribution of mergers to the total ULIRGs luminosity density,
both decline towards higher redshift \citep[e.g.][]{bell05, melbourne05, kartaltepe12, draper12}. 
Others suggested that mergers still play an important role in the most luminous galaxies 
at higher redshift \citep{huang23}. 
According to the morphology quenching theory, 
red galaxies can suppress their star formation 
without exhausting their gas through the presence of bulges that maintain their rotating disks. 
This prevents gas from generating clumps and potentially quenching star formation, 
as supported by several observations \citep[e.g.][]{martig09, guo15}.

In this paper, we present the morphological study of 
a 16$\mu$m selected LIRG-dominated sample at z $\sim$ 1. 
The sample is adopted from  \citetalias{huang21} and  
this is the second paper for this mid-infrared selected sample. 
In the first paper, we performed the SED fitting, 
and managed to separate their AGN and star-forming components. 
In this paper, we will measure their morphological parameters, 
classify their morphological categories, 
study the relations between their morphological parameters, star formation, and AGN activities, 
and explore the evolutionary paths for these mid-infrared selected LIRGs.

This paper is organized in the following order. 
We present our sample in Section \ref{sec:sample}, followed by morphological analyses in Section \ref{sec:morph}. 
In Section~\ref{sec:phy}, we explore the relations between the physical properties and morphological parameters. 
Then, we investigate the possible evolution mechanism of our sample in Section \ref{sec:discussion}, and finally, summarize the conclusions of this paper in Section \ref{sec:summary}. 
The cosmological parameters used throughout this paper are $h \equiv H_0$ [km $s^{-1}$  $Mpc^{-1}$]/100 = 0.7, $\Omega_\Lambda$ = 0.7, $\Omega_M$ = 0.3.

\section{the sample} \label{sec:sample}

The 16$\mu$m selected sample was presented in our first paper in \citetalias{huang21}. 
Here we briefly summarize the selections. 
The parent sample consists of 705 galaxies selected by 
their 16$\mu$m flux from \textit{AKARI}/IRC and \textit{Spitzer}/IRS 
with 0.8 $<$ z $<$ 1.3 in the Extended Groth Strip (EGS), 
and the Great Observatories Origins Deep Survey (GOODS) North and South fields. 
In this redshift range, the PAH features at 6.2, 7.7, and 8.6$\mu$m shift into the observed 16$\mu$m band, 
thus, the \citetalias{huang21} sample selection is chosen to select PAH-dominant galaxies.
We note that mid-infrared bright galaxies could also be due to AGN emission,
but  as demonstrated in \citetalias{huang21}, 
only a fraction (15\%) of our sample have AGN-dominated SEDs. 

Specifically, \citetalias{huang21} carried out a mid-infrared SED analysis for this complete 16$\mu$m selected sample, 
and disentangled 
the star-forming and AGN contributions for each galaxy. 
This sample was then divided into four main types:
 AGN dominated (63), star-forming dominated (248), composite (161) and blue compact (7) galaxies,
see Table 3 in \citetalias{huang21}),
mainly  
based on the 6.2$\mu$m PAH equivalent width (EW), 
and rest-frame 4.5$\mu$m luminosity to 1.6$\mu$m luminosity ratio ($L_{4.5}/L_{1.6}$). 
\citetalias{huang21} also quantified the AGN strength with the 4.5$\mu$m luminosity excess ($L_{4.5}^{Exc}$). 
\citetalias{huang21} showed that most galaxies in this mid-infrared sample 
are in the ``green valley" region (i.e. the transition region in the color-magnitude diagram 
between active star-forming blue galaxies, 
and quiescent red sequence galaxies)
 based on their locations in the color-mass (i.e. rest-frame U-V vs. stellar mass) diagram \citep{salim14}.

In this paper, we only study galaxies in the sample within the CANDLES \citep{koekemoer11candels} coverage, 
in order to obtain morphological information in the observed \textit{HST}/WFC3 F160W band. 
This leaves us a total of  479 galaxies in the CANDLES field in our sample. 
The total infrared luminosities ($L_{IR}$, 8-1000$\mu$m) of our sample are in the range of $10.3<log_{10}(L_{IR}/L_\odot)<12.6$, with most of them (67\%) being LIRGs, and 16 classified as ULIRGs. 
We take advantage of the CANDLES master catalog in \citet[hereafter, F18]{fang18candels} and adopt their redshifts, and stellar masses. 
The stellar masses in the CANDLES catalog were estimated without considering the AGN contamination. 
\citet{huang23} showed that AGNs may contribute up to 25\% to the total luminosity 
in the rest-frame near-infrared bands. 
We argue that the CANDLES stellar mass is not significantly contaminated by AGN, as discussed in Appendix~\ref{appendix:stellar}. 
In this paper, We adopt the star formation rates (SFRs) from \citetalias{huang21}.
{The full list of our sample of 479 galaxies can be found in the online version of the journal (Table \ref{tab:sample}).}

\begin{deluxetable*}{llrcccc}[hbt]
\tablecaption{The 16$\mu$m-selected sample of 479 galaxies in the CANDELS fields \label{tab:sample}}
\tablehead{\colhead{Catalog-ID} & \colhead{R.A.($\degr$)} & \colhead{Dec.($\degr$)} & \colhead{Class} & \colhead{SFR($\frac{\text{M}_{\odot}}{\text{yr}}$)} 
& \colhead{log($\frac{\text{L}_{4.5\mu\text{m}}}{\text{L}_{\odot}}$)} & \colhead{log($\frac{\text{L}_{4.5\mu\text{m,ex}}}{\text{L}_{\odot}}$)}}
\startdata
goodsn-11509 & 189.14378828 & 62.10778178 & SFG & 25.76 & 10.37 & 9.89 \\
goodsn-15112 & 189.18420277 & 62.11301059 & SFG & 7.98 & 9.94 & 9.31 \\
goodsn-13269 & 189.16321833 & 62.11640141 & SFG & 2.99 & 9.10 &  \\
\ldots & \ldots & \ldots & \ldots & \ldots & \ldots & \ldots \\
goodss-1974 & 53.0802579 & -27.9005411 & Composite & 27.78 & 10.32 & 9.76 \\
goodss-24010 & 53.1434268 & -27.9015698 & SFG & 32.4 & 10.04 &  \\
goodss-1928 & 53.1232145 & -27.9013512 & Composite & 9.64 & 9.36 &  \\
\ldots & \ldots & \ldots & \ldots & \ldots & \ldots & \ldots \\
egs-13100714 & 215.1538291 & 52.9661384 & Composite & 11.89 & 9.95 & 9.40 \\
egs-13017614 & 215.1009825 & 52.9280946 & SFG & 61.49 & 10.27 &  \\
egs-13033068 & 215.2823342 & 53.0550711 & AGN & 83.05 & 10.87 & 10.78 \\
\ldots & \ldots & \ldots & \ldots & \ldots & \ldots & \ldots \\
\enddata
\tablecomments{Table \ref{tab:sample} is published in its entirety in the machine-readable format. 
A portion is shown here for guidance regarding its form and content.}
\end{deluxetable*}

\section{morphology of the 16$\mu$m Selected Sample} \label{sec:morph}
Morphological analysis for our sample is carried out on the CANDLES \textit{HST} F160W images. 
All \textit{HST} images used in this paper can be found in Mikulski Archive for Space Telescopes (MAST): \dataset[https://doi.org/10.17909/T94S3X]{https://doi.org/10.17909/T94S3X}. 
Figure~\ref{fig:stamp1} shows the stamp images for all galaxies in our sample. 
We first analyze the distributions of the structural parameters (s{\'e}rsic index and effective radius) 
and find that our sample is dominated by disky galaxies. 
Then we visually identified the galaxy pairs in our sample and galaxies with distorted morphologies utilizing the criteria in \citet{lotz08}. 
This way, we find 72 galaxies in physical pairs and a total of 86 galaxies in our sample that locate in the merger region.

\begin{figure*}[h!]
    \centering
    	\includegraphics[scale=0.5]{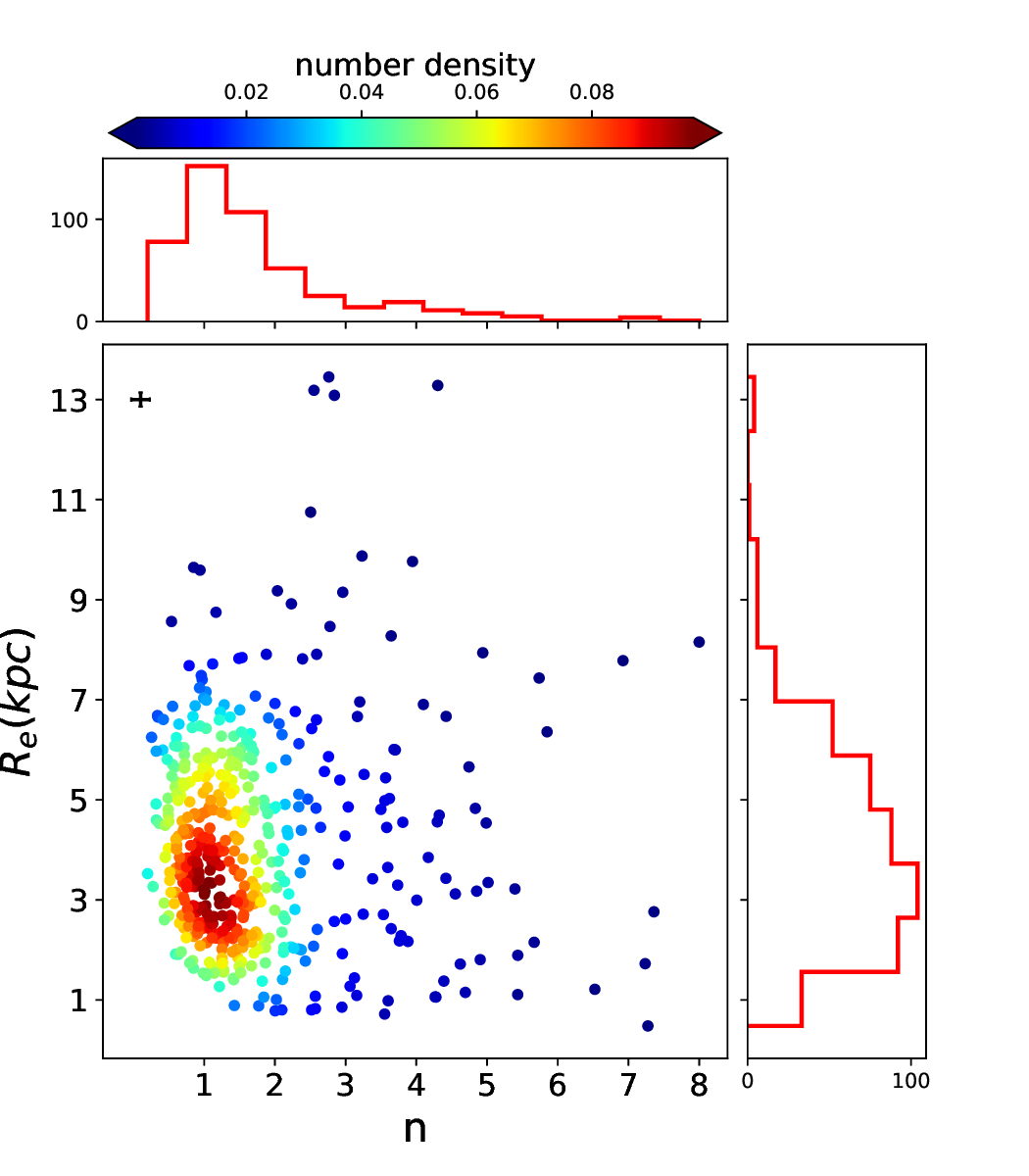}
\includegraphics[scale=0.5]{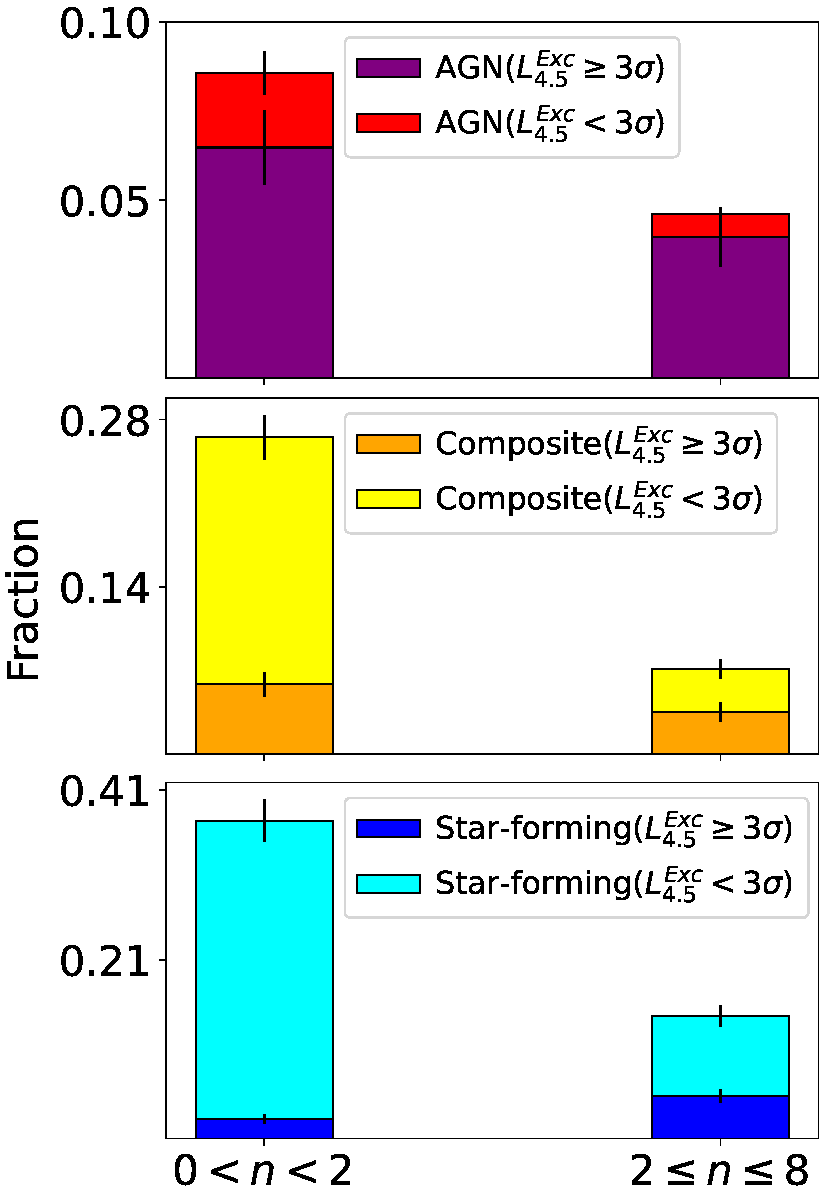}
    \caption{\textit{Left}: Effective Radius $R_e$ and S{\'e}rsic index n distribution for our LIRG-dominated sample, color-coded by number density. 
        The median values of $R_e$ and n are 3.8 kpc and 1.4, respectively. 
        In the upper left of the plot, the crossed error bar represents the 3$\sigma$ typical uncertainty. 
        \textit{Right}: Distribution of n for galaxies with different SED classes, as identified by \citetalias{huang21}. 
        The fractions of AGN dominated galaxies, composite and star-forming dominated galaxies with 2 $\leq n\leq$ 8 are 0.04, 0.07 and 0.14, respectively. 
        To quantify the strength of AGN activity, we use the AGN contribution to the rest-frame 4.5$\mu$m luminosity, denoted as $L_{4.5}^{Exc}$, as in \citetalias{huang21}, where $L_{4.5}^{Exc} > 3\sigma$ represents strong AGN contributions ($\sigma$ is the uncertainty of $L_{4.5}^{Exc}$ from \citetalias{huang21}). 
        We find that most of our galaxies (71\%) are disky galaxies (n $<$ 2) and the fractions of bulge-dominated galaxies (n $\geq2$) are higher in AGN dominated galaxies, compared to star-forming dominated and composite galaxies.}
    \label{fig:morphdistri}
\end{figure*}

To determine the structural parameters of galaxies in our sample, we utilize the s{\'e}rsic index (n) and effective radius ($R_e$) measured by \citet[hereafter, V14]{van14} , also based on \textit{HST} F160W images. 
For the redshifts of our sample, the above structural parameters were measured around rest-frame 8000\AA, 
a wavelength range where the evolved stellar population dominates the emissions. 
The goodness of the GALFIT \citep{peng10galfit} fitting results was categorized into three levels in \citetalias{van14}: good, suspicious, and bad. 
For validation purpose, we measure the half-light radii from the growth curves and find that 
our half-light radii match well with
the majority (98\%) of $R_e$ in \citet{van14}. 
Only nine out of 479 galaxies are outliers, of which six were already classified as 'bad' in \citetalias{van14}. 
We manually perform the GALFIT on these 9 galaxies after masking out nearby objects, 
and find  consistent $R_e$ measurements to our half-light radii results. 
Therefore, for the 9 outliers, we use the new $R_e$ generated by GALFIT results, 
and for the rest of our sample (470/479), we adopt the $R_e$ and n measurements from \citetalias{van14}. 
The left panel of Figure~\ref{fig:morphdistri} displays our sample's $R_e$ and n distribution, 
which have a median $R_e$ of 3.8 kpc and n of 1.4. 
Our sample is dominated by disky galaxies, with over 71\% of galaxies having n less than 2. 
In the following analyses, we consider galaxies with n less than 2 as disky galaxies \citep{bruce14bddecompose}. 
The right panel of Figure~\ref{fig:morphdistri} illustrates the fraction of galaxy types at $n<$2 and $n\geq$2. 
Compared to the fraction of bulge-dominated morphologies in composite (21\%) and star-forming (28\%) galaxies, 
AGN dominated galaxies have a slightly higher fraction (35\%) .

We then try to identify the galaxy  pairs in the sample, adopting a criteria of projected separation $d<30 h^{-1} kpc$,  and a redshift offset $\Delta V < $1000 km\,s$^{-1}$ \citep[e.g.][]{newman12, man16}. 
In addition, we add a stellar mass ratio requirement of greater than 0.1. 
We do not consider pairs with larger stellar mass differences (i.e. mass ratio $<$ 0.1), as minor mergers do not significantly change the morphology of their central galaxy. 
With these criteria, we find 72 galaxies in physical pairs, 
of which 35 galaxies are in spectroscopically-confirmed pairs.

\begin{figure*}[h!]
    \centering
    \includegraphics[scale=0.75]{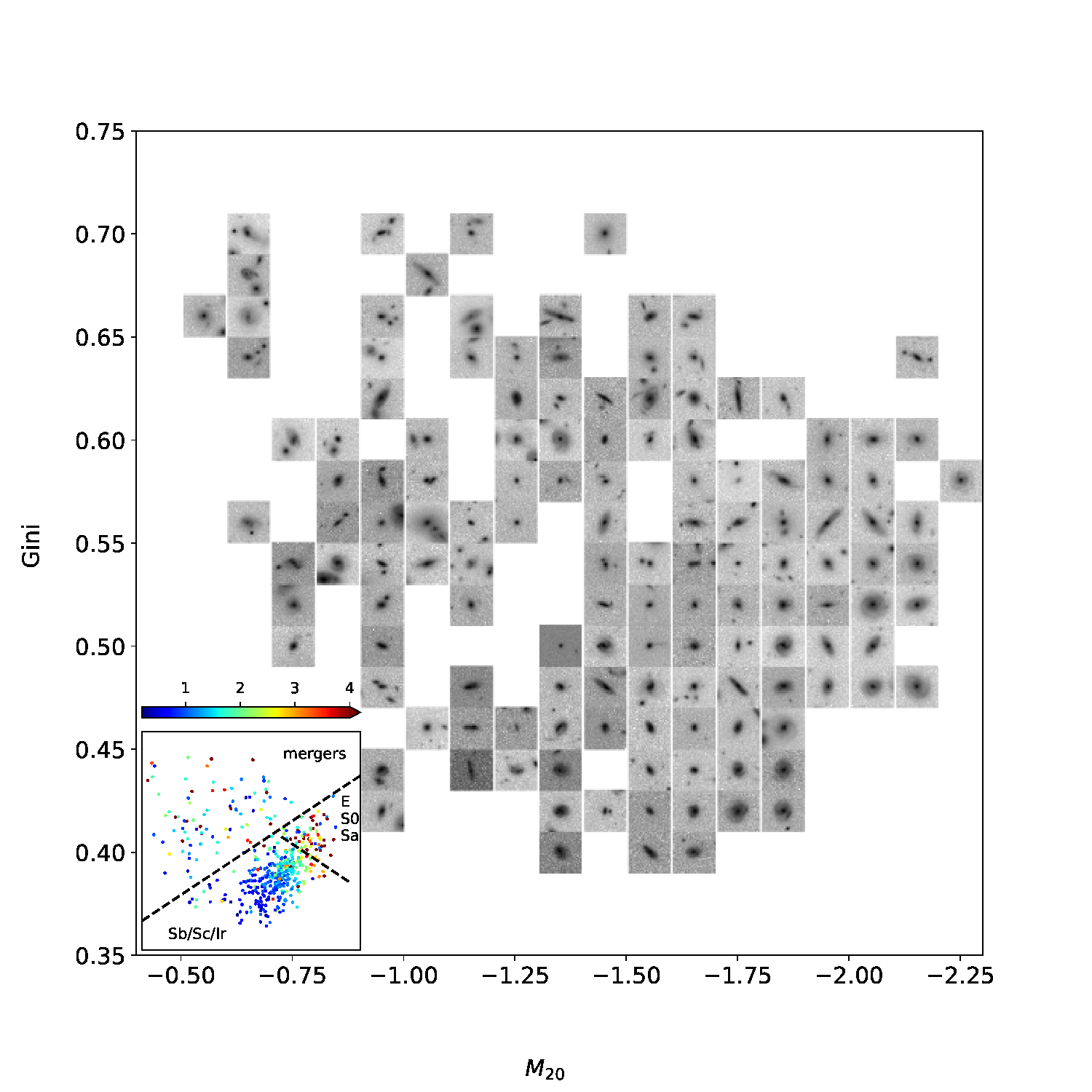}
    \caption{Gini vs. M$_{\text{20}}$ diagram for our LIRG-dominated sample. 
    We show the WFC3 F160W stamp images of the galaxies with median n in each grid. 
    Inset plot is the same plot for our whole sample, color-coded by n.
    We adopt the criteria from \citet{lotz08} to distinguish mergers from E/S0/Sa and Sb-Ir: 
    with G $>$ -0.14 M$_{\text{20}}$ + 0.33 for mergers, 
    G $\leq$ -0.14 M$_{\text{20}}$ + 0.33 and G $>$ 0.14 M$_{\text{20}}$ + 0.80 for E/S0/Sa, 
    and  G $\leq$ -0.14 M$_{\text{20}}$ + 0.33 and G $\leq$ 0.14 M$_{\text{20}}$ + 0.80 for Sb-Ir;
    as shown by the dashed lines in the inset. 
    We find that the Gini vs. M$_{\text{20}}$ distribution of our sample is consistent with previous studies.  
    The  s{\'e}rsic index n shows an  increasing trend along the diagonal line towards smaller M$_{\text{20}}$ and higher Gini in the non-merger region. }
    \label{fig:ginim20}
\end{figure*}

Based on the morphological measurement \citep[e.g.][]{lotz04, lotz08}, M$_{\text{20}}$, 
which is the normalized second order moment of 20\% of the brightest pixels, 
and the Gini index (G), which represents the distribution of the galaxy pixel flux values \citep{lotz08},
we identify galaxies with distorted morphologies in Figure \ref{fig:ginim20}. 
The criteria to separate distorted galaxies from spiral- and bulge- dominant galaxies are adopted from \citet{lotz08}.
Figure~\ref{fig:ginim20} shows the Gini vs. M$_{\text{20}}$ diagram for our sample. 
Based on the criteria of \citet{lotz08}, there are 14 galaxies in pairs located in the merger region,
all of which are in the 35 spectroscopically-confirmed pairs mentioned above. 
The numbers of galaxies located in merger, E/S0/Sa, and Sc/Sd/Ir regions are listed in Table~\ref{tab:gini_m20}. 
Although only 40\% of all pairs can be identified as mergers in the Gini-M$_{\text{20}}$ plot, Figure~\ref{fig:ginim20} shows a clear trend that galaxies located in the merger region are more likely to have distorted morphologies. 
The fraction of galaxies with distorted morphologies that located in the merger region (31\% ) is higher than that located in the non-merger region (5\%). 
Galaxies located in the elliptical region tend to have apparent bulges,  
while in the Sc/Sd/Ir locus, 
galaxies show flatter surface brightness distribution with no or less significant bulges. 
The trends are more evident in the inset plot of Figure~\ref{fig:ginim20}. 
Galaxies in the Sc/Sd/Ir region have lower n values, while galaxies in the elliptical region tend to have higher n (redder color in the inset plot of Figure~\ref{fig:ginim20}). 
This is similar to \citet{wangtao12}, which suggests that elliptical galaxies and disky galaxies with conspicuous bulges tend to exhibit high Gini and low M$_{\text{20}}$, while disky galaxies with less noticeable bulges are more likely to have lower Gini and higher M$_{\text{20}}$. 
Recently, using \textit{JWST} NIRCam images, \citet{Kartaltepe23ginim20ncoded} also find that galaxies at z $=$ 3 - 9 have higher n in the elliptical region, similar to our results at lower z. 
However, an evident proportion of galaxies in the merger region also exhibits higher n in their sample. 
Our sample in the merger region, on the other hand, displays a wide range of n, with only 33\% at n$\geq$2. 
This difference may be due to different sample selections. 
\citet{Kartaltepe23ginim20ncoded} selected galaxies in JWST/NIRCam within the redshift range of z $=$ 3 - 9, which may include quiescent galaxies that have settled down and formed their bulges. 
This may result in the fraction of galaxies with high n in their merger region of the Gini vs. M$_{\text{20}}$ diagram. 
In contrast, based on the 16$\mu$m flux limit we have, the mergers in our sample 
are selected to be dominated
by star-forming galaxies, and most of them 
have not formed prominent bulges yet.

\section{relations between the physical properties and morphologies} \label{sec:phy}
In this section, we examine the relationship between the morphologies of galaxies 
and their physical properties, as well as the possible contributions of  their morphologies to the galaxies' evolutionary process. 
We adopt the star formation rate (SFR) values from \citetalias{huang21}, 
which is mostly converted from the star formation contribution in the rest-frame 8$\mu$m (L$_{8\mu m}^{SFR}$)
after AGN-star formation decomposition.
For the few galaxies (9 in our whole sample) with an AGN dominant 8$\mu$m luminosity (L$_{8\mu m}^{SFR}$ / L$_{8\mu m} < $ 0.5),
\citetalias{huang21} adopted the SFR derived from far-infrared luminosity, which is less affected by the AGN contribution.

\begin{figure*}[h!]
    \centering
    \includegraphics[scale=0.9]{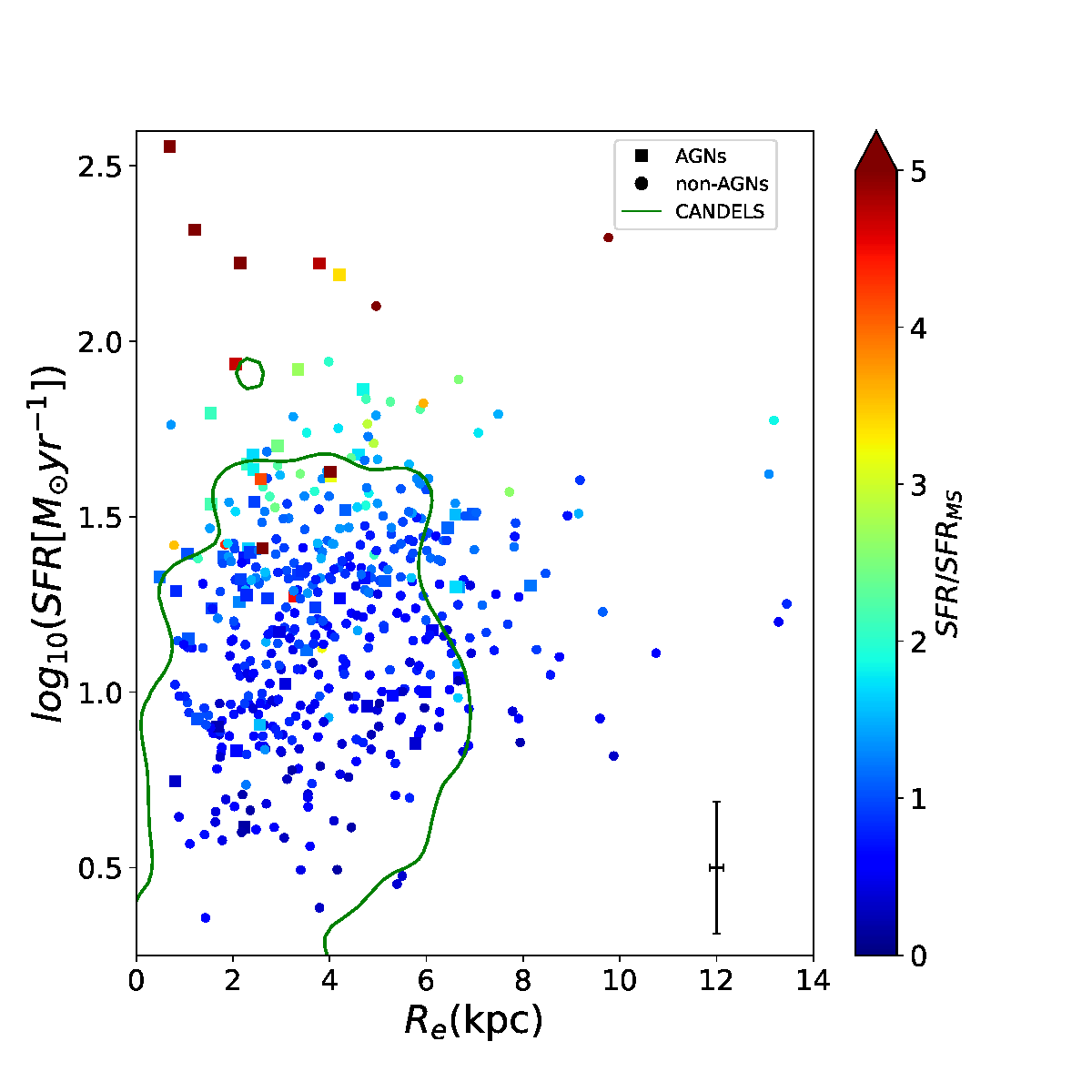}
\vspace{-1cm}
    \caption{The SFR vs. effective radius $R_e$ distribution for our sample. 
    The green contour encompasses 90\% CANDELS galaxies in GOODS-S, GOODS-N and EGS fields with redshifts ranging from 0.8 to 1.3. 
    Squares and dots denote AGN dominated galaxies and non-AGN dominated galaxies in our sample. 
    Each galaxy is color-coded according to its SFR to SFR$_{\text{MS}}$ ratio at the corresponding redshift. 
    The 3$\sigma$ typical uncertainty is shown with the crossed error bar in the bottom right corner. 
    We find no correlation between the SFR and $R_e$ in our sample. 
    Additionally, we find that the majority of galaxies in our sample are normal main sequence galaxies.}
    \label{fig:resfr}
\end{figure*}

First, we investigate the variations in SFRs and $R_e$ for our sample, illustrated in Figure~\ref{fig:resfr}. 
We find no correlation between SFRs and $R_e$ for our LIRG-dominated sample. 
Within the GOODS-S, GOODS-N, and EGS fields, 
90\% of CANDELS galaxies fall inside the green contour as shown in Figure~\ref{fig:resfr}. 
The majority of our sample has high SFRs and occupies the upper part of the CANDELS locus. 
Sixty-seven percent of our sample are LIRGs, as expected since our sample was mostly selected by mid-infrared PAH emissions, which effectively trace SFRs \citep[e.g.][]{peeters04pahsfr, schreiber04pahsfr}. 
As shown in Figure~\ref{fig:resfr}, some AGN dominated galaxies (depicted as squares) exhibit high SFRs.
We note that their SFRs are calculated based on far-infrared luminosities and should not be heavily 
contaminated by AGN contributions. 
\citet{dai12agnfirsfr, madau14SFH} and \citetalias{huang21} found that AGNs contribute little ($< $20\%) to far-infrared luminosities at high star formation levels. 
\citetalias{huang21} showed good agreement between SFR$_{8\mu m}$ and SFR$_{FIR}$, therefore the SFRs for these AGN dominated galaxies are not likely significantly overestimated. 
Given their high absolute values, even with a 50\% AGN contribution, 
the SFR values would only be 0.3 dex lower, still significantly higher than the rest
of the population. 
Our results suggest that strong AGN dominated galaxies can also have high SFRs, 
in other words, 
the AGN dominated galaxies in our sample have not suppressed the star formation activity significantly. 
The luminous cores due to AGN or starbursts in the central region of galaxies may be the reason that resulted in their small size measurements.

\begin{figure*}[h!]
    \centering
        \includegraphics[scale=0.75]{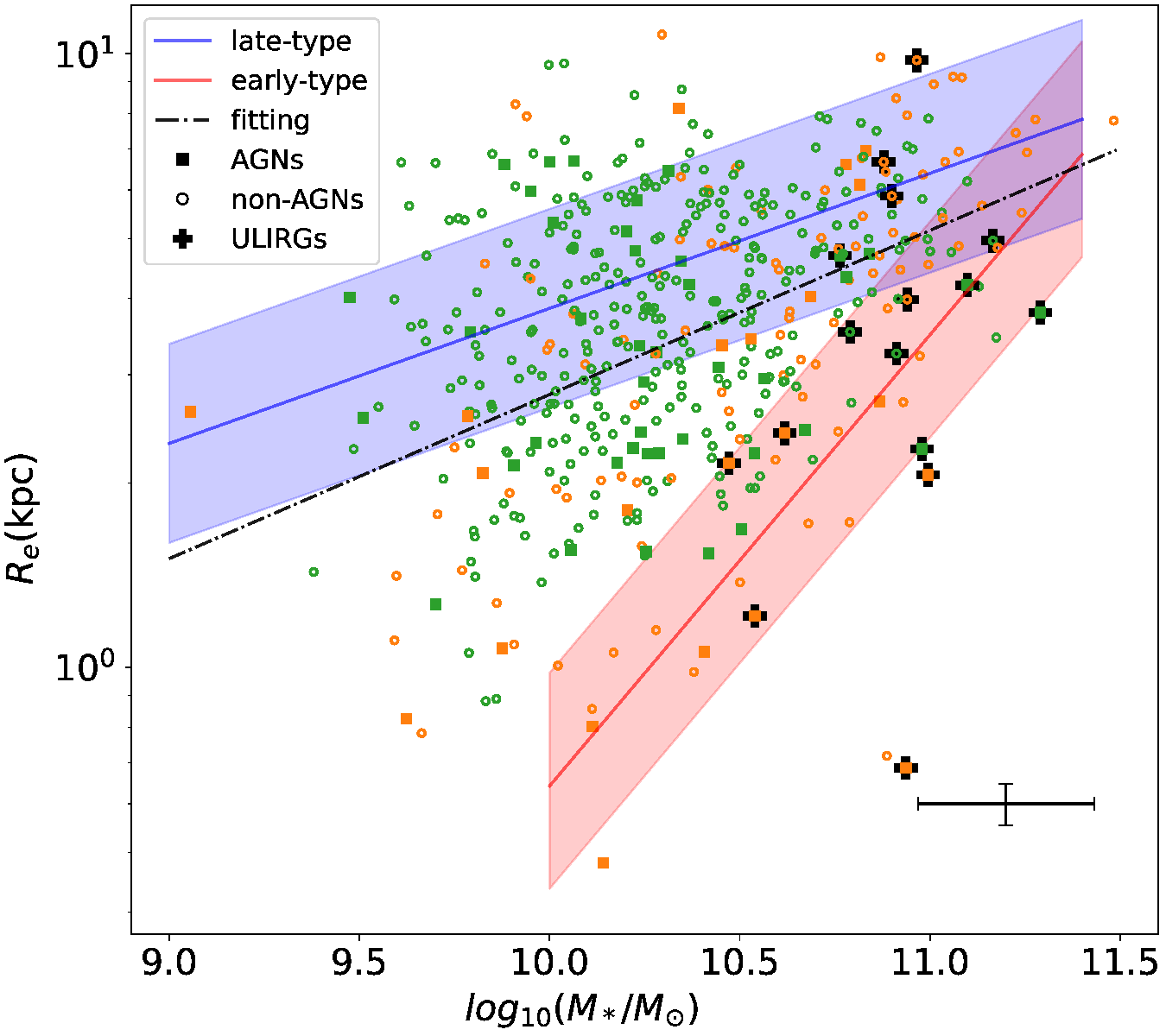}
\caption{Size-mass distribution of our sample. 
    \textit{Symbols}: Squares and open-dots are for AGN dominated galaxies and non-AGN dominated galaxies, respectively, 
    while crosses denotes the ULIRGs in our sample. 
    Blue and red solid lines shaded regions are the size-mass relations and the relevant 3 $\sigma$ regions 
    of late-type and early-type galaxies from \citetalias{van14}. 
    The early-type and late-type galaxies from \citetalias{van14} are defined by their locations in the UVJ diagram. 
    We adopt the \citetalias{van14} correlation to $z=1$, based on linear interpolation of the A and $\alpha$ values at z=0.75 and z=1.25, 
    where A and $\alpha$ are parameters of function $R_{eff}/kpc=A(M_*/5\times 10^{10}M_{\odot})^\alpha$ drawn from 
    Table 1 of \citetalias{van14} (i.e. [log$_{10}A$, $\alpha] =$ [0.74, 0.22] for late-type, and [0.32,0.73] for early-type galaxies). 
    The black dash-dotted line represents the fitting result of our full sample, with fitted [$log_{10}A, \alpha$]=[0.63, 0.26]. 
    Green and orange symbols denote the disky galaxies (n $<$ 2) and bulge-dominated galaxies (n $\geq$ 2) in our sample, respectively. 
    We utilize the uncertainties of stellar masses from \citet{santini15candels}. 
    In the bottom right corner, the crossed error bar represents the 1$\sigma$ typical uncertainty. 
    The majority of our sample (63\%) follow the \citetalias{van14} relation for late-type galaxies (within 3$\sigma$).}
    \label{fig:sizemass}
\end{figure*}

In addition, we also plot the size-mass distribution of the sample in Figure~\ref{fig:sizemass}, adopting similar scales as in \citetalias{van14}. 
The size-mass relations for late-type galaxies and early-type galaxies were found to follow different correlations \citep[e.g.][]{shen03, van14}. 
Late-type galaxies exhibit only a slight increase in size with higher stellar masses, whereas early-type galaxies present a steeper correlation. 
However, at the low-mass end, the sizes of early-type galaxies have no dependence on stellar masses. 
Figure~\ref{fig:sizemass} shows the size-mass distribution of our sample, where the majority of our galaxies (63\%) align with the size-mass relation of late-type galaxies (i.e. within 3$\sigma$ from the relation). 
This confirms that the morphologies of the galaxies in our sample are mostly late-type, 
which is also apparent in the stamp images of Figure~\ref{fig:stamp1}. 
This finding is consistent with \citet[e.g.][]{bell05, melbourne05}, which reported that more than half of LIRGs at z $>$ 0.5 have spiral morphologies. 
We note that the fitted interception for all galaxies in our sample (0.63, black dash-dotted line)
is lower than those of the late-type galaxies(0.74, blue line) in \citetalias{van14}, 
though the fitted slope is comparable (0.26 $vs$ 0.22). 
This difference arises from the fact that our sample, although dominated by disky galaxies ($>$71\% with n $<$ 2),
still includes a significant fraction of galaxies with bulge-dominated (n $\geq$ 2) morphologies.
The smaller R$_e$ of the latter contributes to the lower normalization observed in the fit. 
For the $n \geq $ 2 galaxies in our sample (orange points in Figure~\ref{fig:sizemass}), 
although they do not lie along the \citetalias{van14} size-mass relation for early-type galaxies,
they do tend to be closer to the relation (red line in Figure~\ref{fig:sizemass})
than the $n < $ 2 galaxies in our sample (green points in Figure~\ref{fig:sizemass}). 
The correlation slope is higher for galaxies with $n \geq $ 2 ($\alpha=$0.46) than 
galaxies with $n < $ 2 ($\alpha=$0.20, not plotted in Figure~\ref{fig:sizemass}).
Surprisingly, Figure~\ref{fig:sizemass} also illustrates that many galaxies near the early-type relation are ULIRGs (crosses). 
These ULIRGs have higher stellar masses and align more closely with the size-mass relation of early-type galaxies, 
as shown in Figure~\ref{fig:sizemass}. 
Specifically, 50\% of ULIRGs follow the size-mass relation of early-type galaxies, 
while 25\% follow that of late-type galaxies.
Half of our ULIRGs can be classified as starburst galaxies, 
which is defined as galaxies with a SFR 3 times higher than the main-sequence SFR \citep{elbaz18}. 
Among the galaxies that follow the early-type relation (red line in Figure~\ref{fig:sizemass}), 
AGN-dominated galaxies account for approximately 36\%, 
which is substantially higher than the AGN  fraction in the late-type galaxies (9\%),
and in the full sample (13\%).

\begin{figure*}[h!]
    \centering
    \includegraphics[scale=0.75]{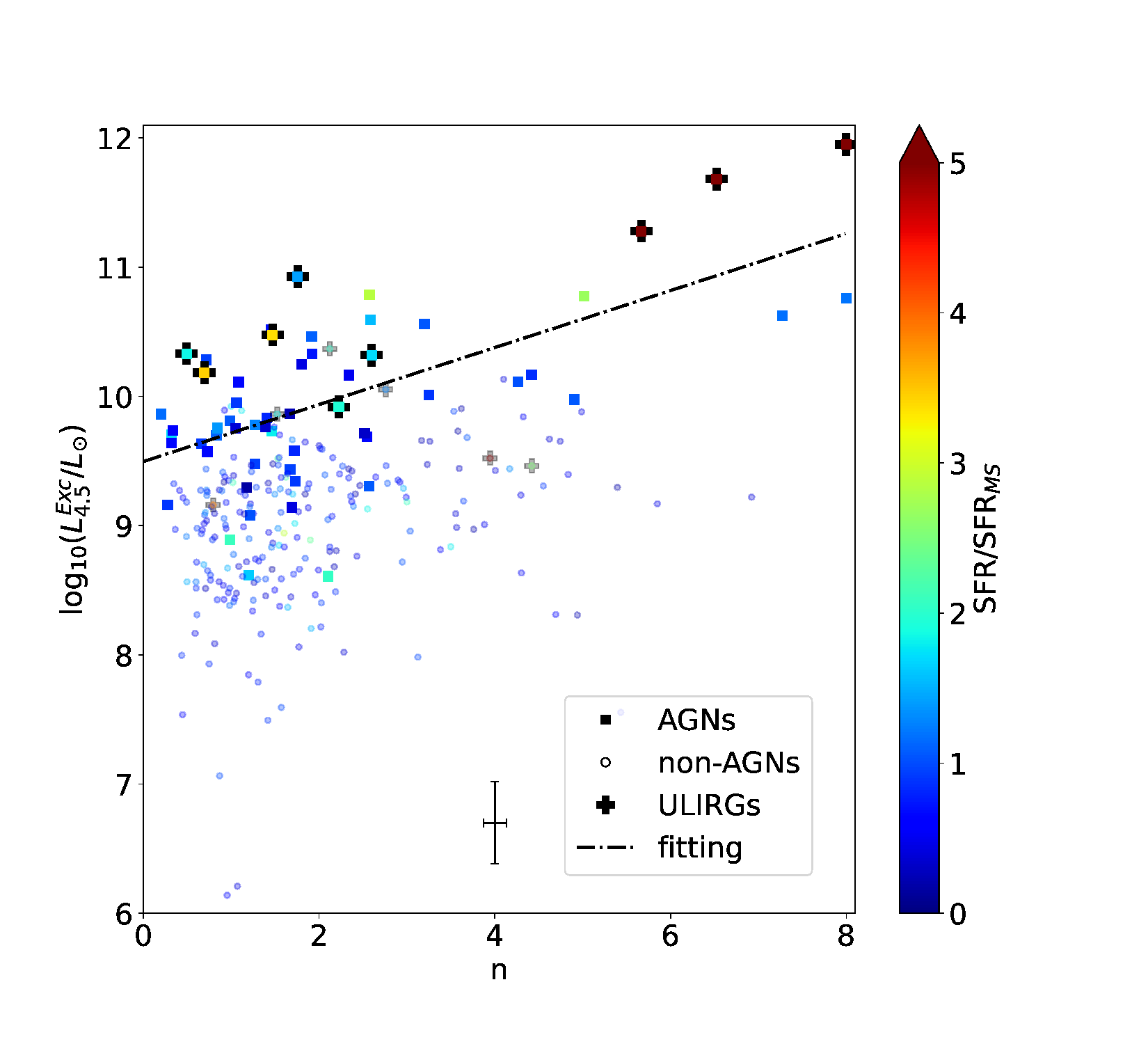}
\vspace{-1.5cm}
    \caption{Distribution of $L_{4.5}^{Exc}$ vs. n for our sample. 
    The symbols are consistent with those in  Figure~\ref{fig:sizemass}.
    We fit the correlation for AGNs in the dot-dashed line, and show the 3$\sigma$ typical uncertainty in the bottom right.
    All galaxies are color-coded by their SFRs to SFR$_{\text{MS}}$ ratios.
    Most AGN dominated galaxies (80\%) have $log_{10}(L_{4.5}^{Exc}/L_{\odot})$ greater than 9.5, while the majority of non-AGN dominated galaxies (86\%) have a value below 9.5.
    Overall, for AGN dominated galaxies, galaxies with higher $L_{4.5}^{Exc}$ also have larger n values. 
    Stronger AGN strength, with higher $L_{4.5}^{Exc}$ and larger n, tend to also have higher SFRs in our sample.}
    \label{fig:agn_n}
\end{figure*}

We investigate whether morphologies are related to AGN activities in Figure~\ref{fig:agn_n}, 
and we find that the AGN strength, $L_{4.5}^{Exc}$,
defined in \citetalias{huang21} as the AGN contribution to the rest-frame 4.5$\mu$m luminosity after AGN-star formation decomposition, 
slightly increases with increasing n (with a slope of 0.22$\pm$0.03, and $p <$ 1e-6). 
In other words, AGN dominated galaxies with stronger nuclear activities
tend to have 
more bulge-dominant morphologies. 
Figure~\ref{fig:agn_n} shows that three AGN dominated galaxies with bulge-dominated morphologies (n$>$5) 
and highest luminosities ($log L_{4.5}^{Exc}>11$) have extremely high SFR/SFR$_{\text{MS}}$ (greater than 5). 
This suggests a possible coevolution between the AGN and star formation activity, consistent with earlier results in \citetalias{huang21} and \citet{dai18}. 
While some AGN dominated galaxies in our sample contain clear bulge components, 
they have not yet quenched their star-forming activities. 
We note that based on the CANDELS rest-frame magnitudes, 
the majority of our galaxies (98\%) are located in the star-forming region of the UVJ diagram, 
though we find no correlation between the 
morphological parameter n, and the galaxies' positions in the UVJ diagram.

\section{discussion}\label{sec:discussion}
\subsection{Galaxy evolution track }
In the merger scenario of galaxy evolution, 
mergers or interactions between galaxies can trigger intense starbursts and AGN activity, 
causing the swift consumption of gas reservoirs and 
resulting in rapid mass growth \citep{hopkins06mergeragnsb}. 
Afterwards, galaxies enter a quiescent stage, and their sizes grow the so-called size evolution \citep{Cassata11sizeevolution}. 
In addition to mergers, \citet[e.g.][]{lefloch05, melbourne05} discussed other mechanisms, such as sufficient gas supply that can support high SFRs at the LIRG level.

We construct the $log_{10}sSFR$ vs. $log_{10}\Sigma_{1.5}$ diagram,
where $\Sigma_{1.5}$ is the stellar mass normalized by effective radius, 
defined as $\Sigma_{1.5}=M_*/R_e^{1.5}$), 
to study the possible evolutionary track of our sample (Figure~\ref{fig:barro}). 
This diagram has been used to diagnose 
if a galaxy evolve through the 'fast' or 'slow' tracks, as shown in \citet[hereafter B13]{barro13}. 
In the slow track,  
most of the star-forming galaxies consume their gas through gas-poor processes and transform 
into typical quiescent galaxies without passing through a compact stage. 
For the fast evolution track,  star-forming galaxies can turn into compact star-forming galaxies through compaction events, such as gas-rich merger and disk instability, which can supply gas to the central region of a galaxy \citep[e.g.][]{dekel13compactevnt, zelotov15compactevent, tacchella16compactevent}. 
In the $log_{10}sSFR$ vs. $log_{10}\Sigma_{1.5}$ diagram, some star-forming galaxies cross the vertical line ($log_{10}\Sigma_{1.5}$=10.4, in \citetalias{barro13} this value is 10.3, in \citet{barro16} this value is 10.4) and evolve into compact star-forming galaxies through this process. 
Compact star-forming galaxies consume their gas during this phase, leading to the quick quenching of their star formation, 
while experiencing minimal changes in their compactness ($log_{10}\Sigma_{1.5}$). 
The resulting compact quiescent galaxies eventually turn into the normal, non-compact quiescent galaxies 
through size growth. 
In \citetalias{barro13}, it is suggested that galaxies may evolve following either a slow or fast track, with different mechanisms dominating galaxy evolution at different redshifts. 
The fast track may dominate at z = 3 $-$ 2, 
while the slow track dominates at z $\lesssim$ 2. 
Our $z\sim1$ sample follows the slow-track in Figure~\ref{fig:barro}, which is consistent with \citetalias{barro13}.

Most galaxies in our sample are located in the non-compact region, 
to the left side of the vertical lines ($log_{10}\Sigma_{1.5}=10.4$), and inside the green contour of CANDELS galaxies. 
These galaxies follow the general evolutionary trend of typical star-forming galaxies in this redshift range (i.e., the slow track of \citetalias{barro13}), 
suggesting that they may not have undergone gas-rich processes. 
In addition, eight sources locate in the compact region of Figure~\ref{fig:barro}. 
Six of these 8 compact sources are bulge-dominated galaxies with $n>2.2$ and $R_e<2.1$ kpc (median $R_e$ of our whole sample is $3.8$ kpc), which are not extended sources and have no distorted signature. 
Indeed, six AGN dominated galaxies of the 8 compact galaxies have high X-ray luminosities ($log_{10}L_{X}>10^{42.1} erg/s$ ) and significant AGN strength ($log_{10}(L_{4.5}^{Exc}/L_{\odot})>10^{9.9}$). 
This is consistent with the second scenario, 
suggesting that they are in transition between star-forming galaxies and quasars. 
However, we cannot specify the possible physical processes behind the compactness for the other two non-AGN dominated galaxies. 
Their uncertainties of $log_{10}sSFR$ and $log_{10}\Sigma_{1.5}$, on the other hand, are quite large, 
making it hard to draw any firm conclusions.

Figure~\ref{fig:barro} shows that most of our AGN dominated galaxies are located in the star-forming region 
and have a slightly higher $log_{10}sSFR$ (median $log_{10}sSFR[Gyr^{-1}]$ of AGN dominated galaxies is 0.1, 
that of non-AGN dominated galaxies is -0.1) than the whole sample. 
In our sample, among the galaxies with the top 10 highest SFRs, 
7 are AGN-dominant,
indicating no significant evidence of AGN quenching, 
at least not yet in our LIRG-dominated sample.

\subsection{Behavior of ULIRGs in our sample} \label{sec:sbs}
Upon analyzing the morphologies of our sample, 
we show several observational relations between $R_e$, n, SFRs, stellar masses, and AGN strengths. 
The majority of our sample exhibits characteristics typical of main-sequence star-forming galaxies at high redshift. 
Specifically, they are disky galaxies with a median n value of 1.4, 
inside the CANDELS locus for high SFR galaxies (Figure~\ref{fig:barro}, green contour). 
It is likely that these galaxies evolve via secular evolution (aka slow track), 
rather than through more intense process such as gas-rich mergers \citep{draper12, larson16}. 
The ULIRGs of our sample are more compact (median compactness higher by at least 0.5 dex) 
than other galaxies at similar sSFR, 
indicating possible transition to compact, star-forming galaxies,
consistent with the GOALs finding \citep{psychogyios16} at much lower redshift ($z<$0.088). 
In Figures~\ref{fig:sizemass}, \ref{fig:agn_n}, \ref{fig:barro}, 
ULIRGs often appear as outliers.
Out of the 16 ULIRGs in our sample, 9 are AGN dominated galaxies. 
Here, we explore the possible reasons for these differences 
and how their morphologies may regulate their evolution.

According to \citet{sanders88}, mergers can trigger both starbursts and AGNs. 
Among the 16 ULIRGs, six appear to be physical pairs (See Section~\ref{sec:morph}), 
including 3 bulge-dominated galaxies with n $\geq$ 2. 
Another two ULIRGs are located in the merger region in Gini vs. M$_{\text{20}}$ diagram, 
which suggests that they may have been through merging process. 
The high SFRs for these ULIRGs may be related to the enhancement of star formation activities
triggered by galaxy interactions \citep[e.g.][]{kennicutt87, Lambas03sfrenhance, woods06sfrenhance, licheng08sfrenhance}. 
Among the remaining eight ULIRGs, 
three galaxies have strong AGN strength of $L_{4.5}^{Exc}>10^{11.2}L_{\odot}$ 
and high sersic index (n $>$ 5.5), 
with \textit{HST} F160W FWHM (full width at half maximum) at most $<$1.7 pixel bigger than 
that of the PSF (point spread function, 3.3 pixels). 
Another three ULIRGs also have apparent bulge in \textit{HST} F160W images and n $>$ 2. 
These 6 ULIRGs with apparent bulges may be in transition to have a dominant core. 
In simulations and observations, gas-rich mergers \citep[e.g.][]{hopkins06mergeragnsb, younger09mergeragnsb, donley10mergeragnsb} and disk instability \citep[e.g.][]{bower06diskinstability,hickox09evolutionprocesses,menci14diskinstability} can break the stable state of galaxies 
and fuel gas to both nuclear and star forming activities. 
Additional information such as rotational velocity
is needed to distinguish between the gas-rich merger or disk instability
scenarios for these 6 ULIRGs \citep[e.g.][]{mo98diskinstability, cole00diskinstability, bower06diskinstability}. 
The remaining two ULIRGs are star-forming dominated galaxies with n of 1.5 and 1.7, respectively. 
We do not identify any significant merging features (such as double cores or tidal tails) 
in their \textit{HST} F160W images. 
These two galaxies could still be post-mergers, 
or have a surge of external gas supply \citep[e.g.][]{guijarro22}.

At similar sSFRs (in each $log_{10}sSFR$ bin with a bin size of 0.5), 
the median values of compactness ($log_{10}\Sigma_{1.5}$) for 
ULIRGs in our sample are always at least 0.5 dex higher (i.e. more compact)
than non-ULIRGs. 

\section{summary} \label{sec:summary}
\begin{figure*}[h!]
    \centering
    \includegraphics[scale=0.75]{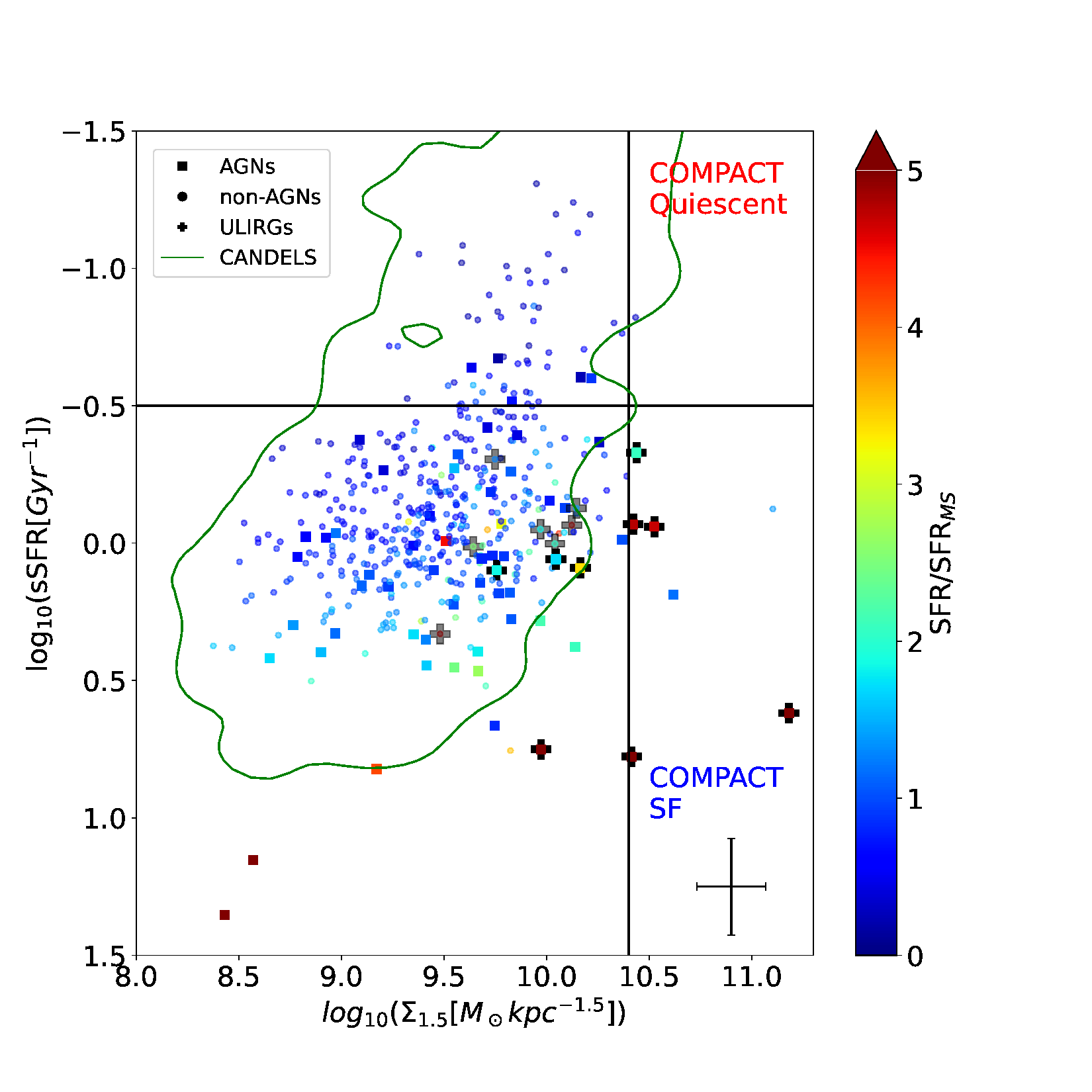}
\vspace{-1.5cm}
    \caption{The distribution of the specific SFR (log(sSFR[Gyr$^{-1}$])
    and the compactness (log($\Sigma_{1.5}$[M$_{\odot}$kpc$^{-1.5}$]) of our sample, where $\Sigma_{1.5}$ is defined as $M_*/R_e^{1.5}$ \citepalias{barro13}. 
    The horizontal line separates quiescent galaxies (above) from star-forming galaxies (below), and the right side of the vertical line represents that galaxies may have compact morphologies. 
    Symbols are the same as in Figure~\ref{fig:agn_n}, and the 1$\sigma$ typical uncertainty is denoted by a crossed error bar at the bottom of the plot.
    The green contour shows the distribution of 90\% for galaxies in the CANDELS GOODS-S, GOODS-N and EGS fields with redshifts ranging from 0.8 to 1.3.
    The distribution of our sample agrees in general with the secular evolution track from \citetalias{barro13}.}
    \label{fig:barro}
\end{figure*}

We conduct a morphological analysis of a 16$\mu$m PAH-selected LIRG-dominated sample from \citet{huang21} in the 
three CANDELS fiels (GOODS-S, GOODS-N, EGS), and investigate the relations between their morphological and physical properties. 
Our findings are as follows.
\begin{itemize}
\item{Based on the \textit{HST/WFC3} F160W images, galaxies in our sample exhibit a variety of morphologies regardless of their SED classes. 
However, our sample has a median n value of 1.4, and the majority of our galaxies (71\%) have n $<$ 2, 
suggesting a dominance of disky galaxies. 
About  35\%  of the AGN dominated galaxies appear to be bulge-dominant,
while that percentage is 28\%  for star-forming and 21\% for composite galaxies.}

\item{The Gini vs. M$_{\text{20}}$ distribution of our sample is in good agreement with previous diagnosis, where galaxies with apparent bulges and larger $n$ are located in the E/S0/Sa region, while those with extended disk components and lower $n$ are more inclined to be in the Sc/Sd/Ir locus. 
Galaxies in the merger region tend to have higher distorted morphologies fraction (31\%) than in the non-merger region (5\%),
while the median n for mergers ($n=$1.3) is quite similar to the full sample ($n=$1.4)}.

\item{The majority of this 16$\mu$m-selected sample are main sequence galaxies, 
though the 16 ULIRGs show more compact morphologies (at least 0.5 dex higher compactness) and significant AGN strengths ($log_{10}(L_{4.5}^{Exc}/L_{\odot})>9$). 
These ULIRGs behave differently in almost all our diagrams (size vs. mass, $log_{10}L_{4.5}^{Exc}$ vs. n, sSFR vs. compactness, 
Figure~\ref{fig:sizemass}, \ref{fig:agn_n}, \ref{fig:barro}). 
Among the 16 ULIRGs, 
six are physical galaxy pairs, 
two have distorted morphologies, 
five have n $>$ 2, 
and the remaining three star-forming dominated galaxies have no obvious merging signature.}

\item{The majority of our sample (63\%) follow the size-mass relation of late-type galaxies within 3$\sigma$. 
In our early-type galaxies, 
there is a higher fraction (36\%) of AGN dominated galaxies
than in the full sample (13\%). 
In addition, we find that galaxies with increasing AGN strength become more bulge-dominant.
The most luminous AGNs in our sample also have the highest SFRs, 
indicating possible coevolution and a lack of signs of AGN-quenching.}

\item{ Most galaxies in our sample follow the secular evolution track (aka `slow' track), 
based on their sSFR and compactness distribution.
While seven galaxies lie in the compact star-forming region, 
almost all (6 out of 7) of them are AGN dominated galaxies, indicating different evolution mechanism.}
\end{itemize}

\section*{Acknowledgement}
{This work is sponsored by the National Key R\& D Program of China (MOST) for grant No.\ 2022YFA1605300, 
the National Nature Science Foundation of China (NSFC) grants No.\ 12273051 and \ 11933003. 
Support for this work is also partly provided by the CASSACA. 
This work is based on observations taken by the CANDELS Multi-Cycle Treasury Program with the NASA/ESA HST, which is operated by the Association of Universities for Research in Astronomy, Inc., under NASA contract NAS5-26555.}

\newpage

\begin{table*}[h!]
    \centering
    \begin{tabular}{c c c c}
        \hline
        \hline
         & Merger & E/S0/Sa & Sc/Sd/Ir \\
        \hline
        whole sample & 86 & 64 & 329 \\
        galaxies in physical pairs & 14 & 6 & 15 \\
        \hline
    \end{tabular}
    \caption{The numbers of our galaxies located in the Merger, E/S0/Sa, and Sc/Sd/Ir regions. 
    We also list the number of galaxies that are spectroscopically-confirmed physical pairs.}
    \label{tab:gini_m20}
\end{table*}

\begin{figure*}[h!]
    \centering
    \vspace{-0.5cm}
    \includegraphics[scale=0.25]{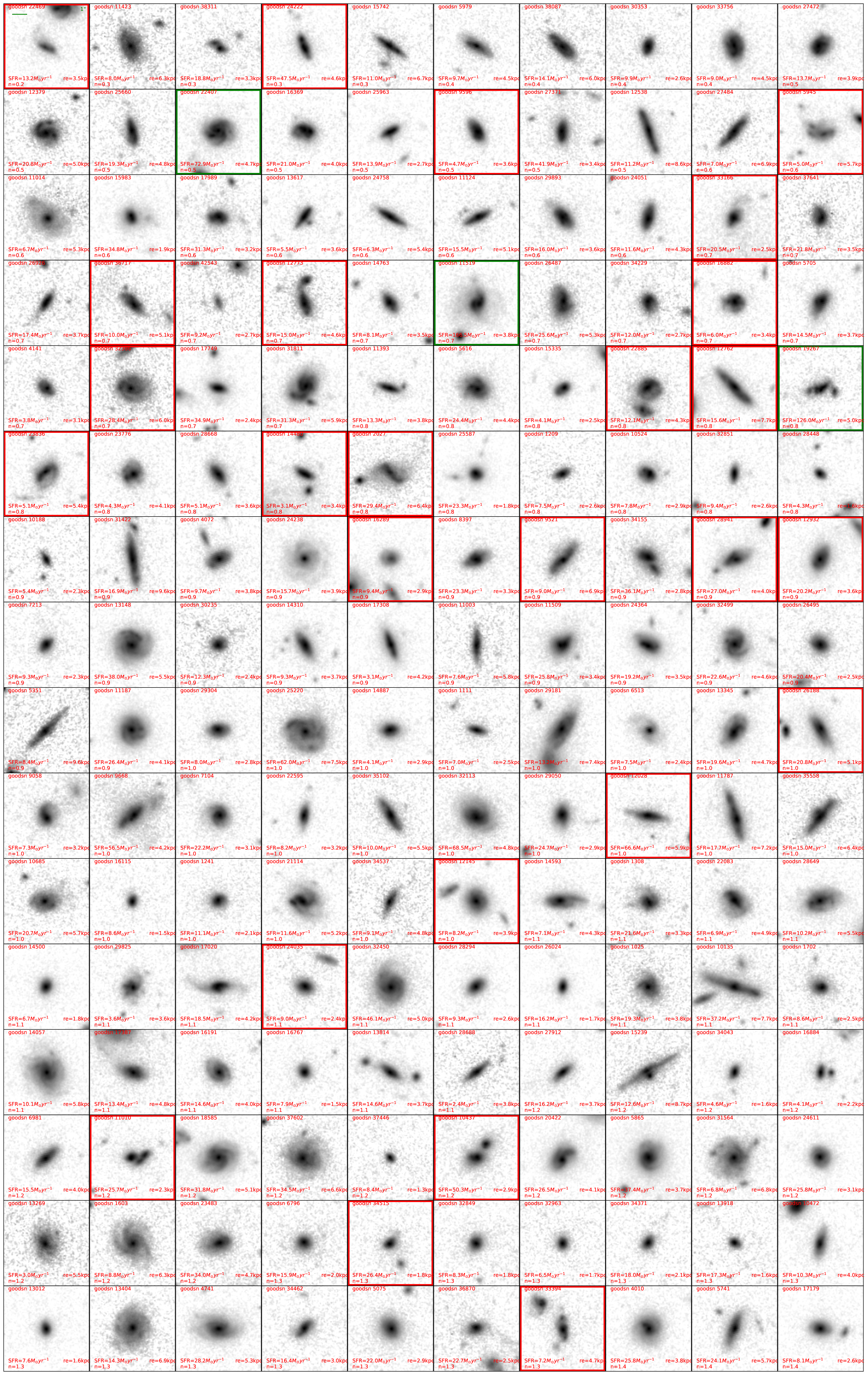}
    \vspace{-3cm}
    \caption{\textit{HST} WFC3 F160W stamp images for our LIRG-dominated sample. 
    ULIRGs are highlighted with green borders. 
    Red rectangles mark the physical pairs mentioned in Section~\ref{sec:morph}. 
    Note that often times only one member of the pair system is  shown in the cutout images. 
    Various morphologies exist in our sample.}
    \label{fig:stamp1}
\end{figure*}

\begin{figure*}[h!]
   \centering
   \vspace{-0.5cm}
   \includegraphics[scale=0.25]{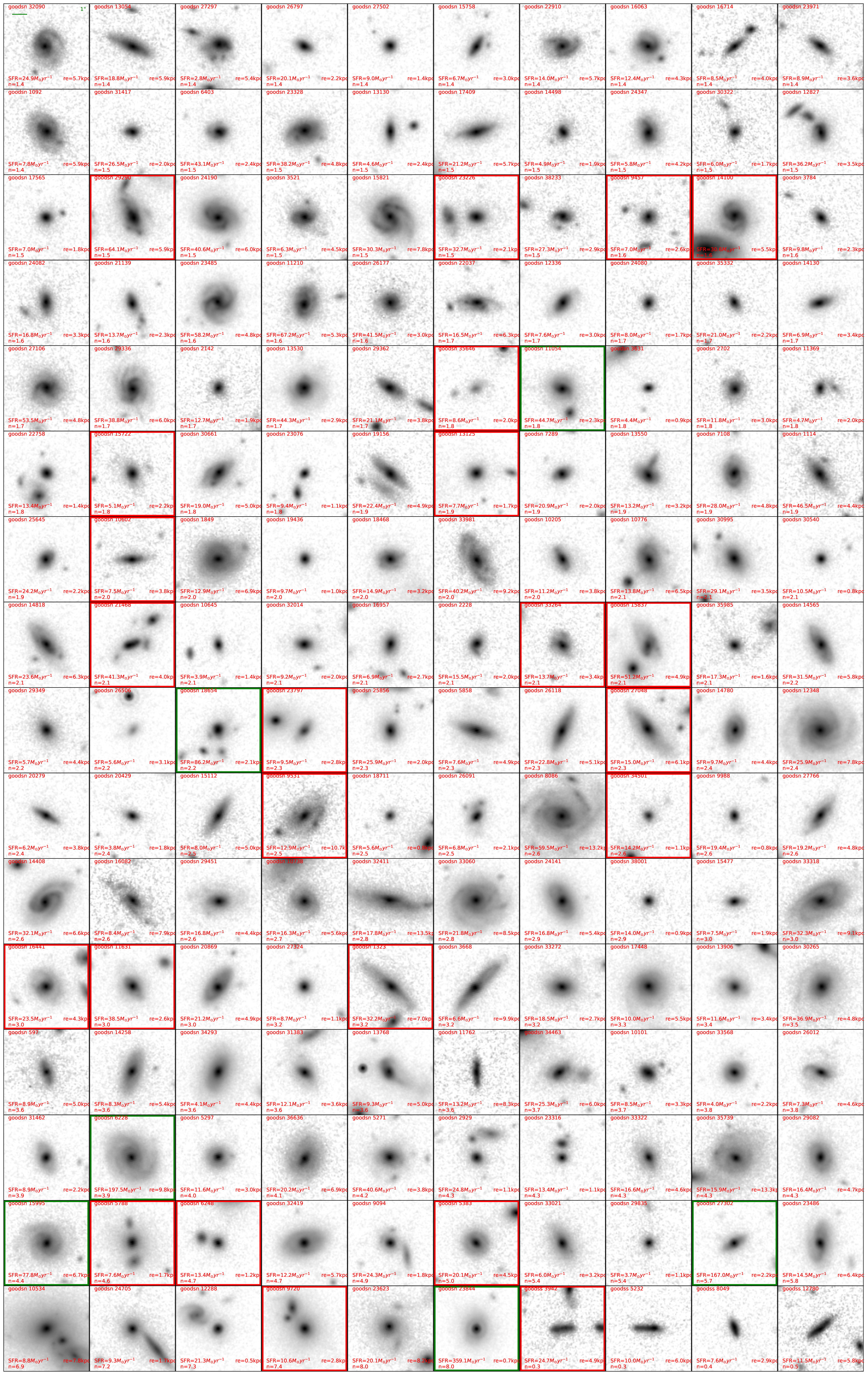}
   \vspace{-3cm}
   \contcaption{Continued.}
   \label{fig:stamp2}
\end{figure*}

\begin{figure*}[h!]
    \centering
    \vspace{-0.5cm}
    \includegraphics[scale=0.25]{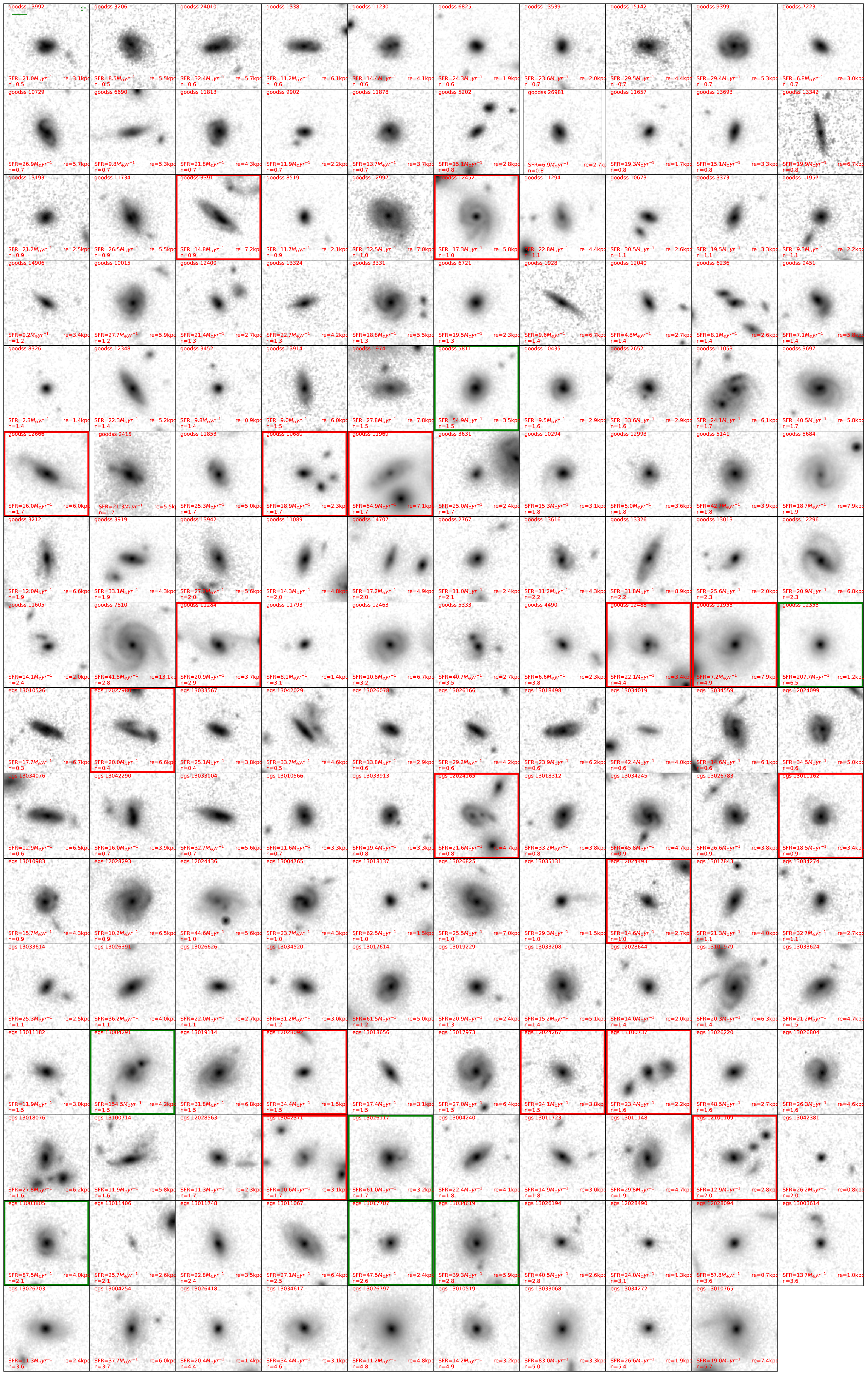}
    \vspace{-3cm}
    \contcaption{Continued.}
    \label{fig:stamp3}
\end{figure*}

\appendix
\section{Stellar mass correction}
\label{appendix:stellar}
\renewcommand{\thefigure}{A\arabic{figure}}
\renewcommand{\thetable}{A\arabic{table}}
\setcounter{figure}{0}
\setcounter{table}{0}

\begin{figure*}[h!]
    \centering
    \includegraphics[scale=0.7]{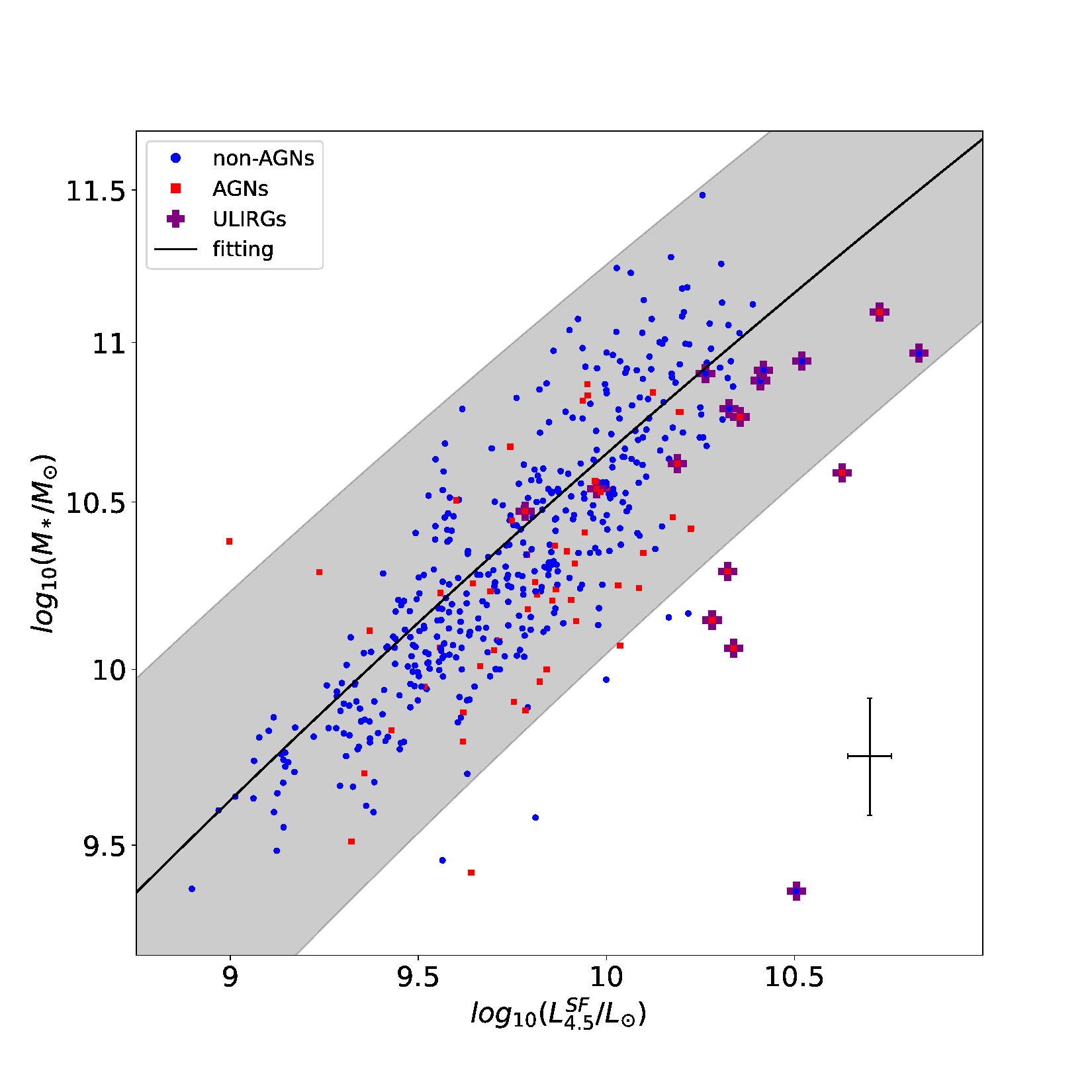}
\vspace{-1.2cm}
    \caption{Stellar masses vs. $L_{4.5}^{SF}$ correlation for AGN dominated galaxies and non-AGN dominated galaxies in our sample. 
    The red squares represent AGN dominated galaxies, the blue dots are non-AGN dominated galaxies, and the purple crosses are ULIRGs mentioned in Section \ref{sec:sbs}. 
    We show the 1$\sigma$ typical uncertainty in the crossed error bar at the lower right side. 
    The $L_{4.5}^{SF}$ is from \citetalias{huang21}, representing 1.6$\mu$m luminosities. 
    We plot the correlation for non-AGN dominated galaxies as the black line, with the shaded area indicating the 3$\sigma$ limit. 
    The stellar masses of AGN dominated galaxies generally follow the black solid line correlation. 
    However, for galaxies outside the 3$\sigma$ limit, we correct the stellar masses accordingly.}
    \label{fig:mass_corr}
\end{figure*}

Stellar masses were derived from CANDELS data, which used Near Ultraviolet to near-infrared multi-wavelength data for SED fitting \citep{santini15candels, mobasher15candels, fang18candels}. 
The presence of AGNs in about 13\% of galaxies in our sample may lead to an over-estimate of the stellar masses.
\citetalias{huang21} found a nearly constant $L_{4.5\mu\,m}/L_{1.6\mu\,m}$ for elliptical galaxies, suggesting a strong correlation between 4.5$\mu$m emission from stellar photospheres and the 1.6$\mu$m stellar bump. 
To estimate the contamination of AGNs to stellar masses, we used $L_{4.5}^{SF}$ ($L_{4.5}^{SF}=L_{4.5}-L_{4.5}^{Exc}$) from \citetalias{huang21} and the stellar masses from CANDELS. 
Figure~\ref{fig:mass_corr} shows the $M_*-L_{4.5}^{SF}$ correlation for our sample, demonstrating that AGN dominated galaxies span the same range of both stellar mass and $L_{4.5}^{SF}$ as the non-AGN dominated galaxies. 
We use non-AGN dominated galaxies in our sample to fit the correlation between stellar masses and $L_{4.5}^{SF}$,
and derive the correlation as $log_{10}(M_*/M_{\odot})=klog_{10}(L_{4.5}^{SF}/L_{\odot})+b$. 
The fitting result yields values of k$=$ 1.02 and b$=$0.39. 
The stellar masses of the galaxies which are outside the $\pm$3$\sigma$ range of the correlation are corrected and shown in Table~\ref{tab:mass_corr}. 
These outliers have a median value of 0.6 dex of the differences between corrected and uncorrected stellar masses. 
We also check the correlation between rest-frame 1.6$\mu$m luminosities and stellar masses of non-AGN dominated galaxies in our sample. 
The results of these two correlations ($L_{1.6\mu m}$ vs. $M_*$, $L_{4.5}^{SF}$ vs. $M_*$) 
are consistent, except for one galaxy (brown square at the bottom left of Figure~\ref{fig:barro}). 
Based on the consistencies, we conclude that the results of this paper are not significantly influenced 
by the uncertainties in the stellar mass estimates of AGN dominated galaxies.

\begin{table*}[h!]
    \centering
    \begin{tabular}{c c c}
        \hline
        \hline
        ID & CANDELS $log_{10}(M_*/M_{\odot})$ & $log_{10}(M_*/M_{\odot})$ after correction \\
        \hline
        GOODS-N 9988 & 10.38 & 9.62 \\
        GOODS-N 10135 & 10.15 & 10.6 \\
        GOODS-N 11054 & 10.28 & 10.84 \\
        GOODS-N 11393 & 9.45 & 10.20 \\
        GOODS-N 11519 & 10.59 & 11.28 \\
        GOODS-N 12028 & 10.16 & 10.87 \\
        GOODS-N 18654 & 10.06 & 10.72 \\
        GOODS-N 19267 & 9.37 & 9.886 \\
        GOODS-N 21468 & 10.06 & 10.68 \\
        GOODS-N 23844 & 10.14 & 11.21 \\
        GOODS-N 34515 & 9.57 & 10.21 \\
        GOODS-N 38311 & 9.42 & 10.28 \\
        GOODS-S 2652 & 9.97 & 10.46 \\
        \hline
    \end{tabular}
    \caption{The results of mass correction. The $log_{10}(M_*/M_{\odot})$ after correction use the correlation between stellar masses and $log_{10}(L_{4.5}^{SF}/L_{\odot})$.}
    \label{tab:mass_corr}
\end{table*}

\bibliography{reference.bib}
\end{CJK*}
\end{document}